\documentclass[fleqn,twoside]{article}
\usepackage{amsmath}
\usepackage[headings]{espcrc1}
\usepackage{amssymb}
\usepackage{epsfig}

\readRCS
$Id: espcrc2.tex,v 1.2 2004/02/24 11:22:11 spepping Exp $
\usepackage{graphicx}
\usepackage[figuresright]{rotating}

\newcommand{\ra}[1]{\renewcommand{\arraystretch}{#1}}
\newcommand{\avk}{\langle k \rangle}
\newcommand{\average}[1]{\langle k^{#1} \rangle}

\title{Percolation on bipartite scale-free networks}

\author{H. Hooyberghs\address[KUL]{Instituut voor Theoretische Fysica, Katholieke Universiteit Leuven,\\
        Celestijnenlaan 200D, B-3001 Heverlee, Belgium}\thanks{E-mail: hans.hooyberghs@fys.kuleuven.be},
        B. Van Schaeybroeck\addressmark[KUL]$^{,}$\address[KMI]{Koninklijk Meteorologisch Instituut (KMI),\\ Ringlaan 3, B-1180 Brussels, Belgium},
       J. O. Indekeu\addressmark[KUL]}

\runtitle{Percolation on bipartite scale-free networks}
\runauthor{H. Hooyberghs}

\begin{document}

% typeset front matter (including abstract)
\maketitle
\begin{abstract}
Recent studies introduced biased (degree-dependent) edge percolation as a model for failures in real-life systems. In this work, such process is applied to networks consisting of two types of nodes with edges running only between nodes of unlike type. Such {\it bipartite} graphs appear in many social networks, for instance in affiliation networks and in sexual contact networks in which both types of nodes show the scale-free characteristic for the degree distribution. During the depreciation process, an edge between nodes with degrees $k$ and $q$ is retained with probability proportional to $(kq)^{-\alpha}$, where $\alpha$ is positive so that links between hubs are more prone to failure. The removal process is studied analytically by introducing a generating functions theory. We deduce exact self-consistent equations describing the system at a macroscopic level and discuss the percolation transition. Critical exponents  are obtained by exploiting the Fortuin-Kasteleyn construction which provides a link between our model and a limit of the Potts model.
\vspace{1pc}
\end{abstract}

\section{INTRODUCTION}
Scale-free self-similar structures occupy a prominent place among Nihat Berker's research tools. He pioneered an interesting class of hierarchical sets, commonly called ``Berker lattices" \cite{Ostlund,Domany} on which a family of renormalization group transformations is satisfied exactly. Further, the geometrical critical phenomenon of percolation \cite{BerkPerc} and, recently, the cooperative behavior of interacting degrees of freedom on random scale-free networks \cite{BerkSF} has enjoyed his thorough attention. In this paper, we discuss a subtle variant of a percolation process on scale-free graphs of bipartite structure.

In recent decades, a detailed analysis of many real-life structures revealed the existence of a common property. Studies of social, technological and biological networks have indicated that the probability distribution $P(k)$ of the degree $k$ of a node, i.e., the number of links attached to the node, follows a decreasing power law $P(k)\propto k ^{-\gamma}$ for large values of $k$. Notable examples of such so-called scale-free networks are the network of co-authors of scientific articles, protein interaction networks, power grids and transportation networks. An accurate determination of the topological exponent (or ``degree exponent") $\gamma$ in real-life networks is far from easy, but most studies indicate a value of $\gamma$ in the range $2 < \gamma < 3.5$~\cite{newmanrev}.

Most properties of scale-free networks are determined by the existence of a small number of nodes which have a large number of links, commonly called hubs. An important consequence of the existence of these hubs is the resilience of the network against random removal of links or nodes. Such destruction processes are studied in the percolation problem, which mostly focusses on the behavior of the largest cluster in the network if links or nodes are removed. When all links and nodes are present in a (infinite) network, a finite fraction of nodes will belong to the largest or ``giant" cluster. However, when more and more links are removed, a point will be reached at which the fraction of nodes contained in the giant cluster vanishes. This point, which shows all characteristics of a second-order phase transition, is called the percolation transition. If the breakdown of the giant cluster occurs when a {\it finite} fraction of links is still present in the network as a whole, the network is called \textit{fragile}. However, if after the removal of an arbitrary fraction of links the giant cluster is still present, the network is called \textit{robust}.

Studies of percolation on scale-free networks have revealed that the behavior of the giant cluster depends on the topological exponent $\gamma$~\cite{cohen,cohen39}. If $\gamma >3$, the giant cluster breaks down at a finite fraction of links, thus the network is fragile. However, when $\gamma<3$, the network is robust. Most real-life networks are thus robust against random removal of links.
Among the numerous applications, the percolation problem has for instance been studied as a model for failures in the Internet structure and in terrorist networks~\cite{cohen}, or as a model for virus spreading on social networks~\cite{kenah}.

Not all networks can be described using a single type of vertex. For instance, the network of hetero-sexual contacts can best be modeled by two types of vertices, men and women, where links always run between nodes of unlike type. Such networks are called \textit{bipartite networks}. They appear in many social structures, for instance in  affiliation networks, i.e., networks of individuals joined by common membership of communities. The two types of vertices then represent the individuals and the groups, while the links between them indicate group membership~\cite{newmanrev}. One could for instance model the research interests of physicists: the first type of nodes  then represents the researchers, while the second type of nodes consists of the research topics (for instance classified according to the PACS).   Most studies concerning bipartite networks focussed on the architecture and the building process of networks. This is for example the case for sexual-contact networks~\cite{ergun}, for listening habits and music genres~\cite{lambiotte} or for general affiliation networks~\cite{ramasco,peruani}. Among these studies, it was revealed that in many bipartite networks, the degree distribution of both types of nodes shows the scale-free characteristic, for instance in the hetero-sexual contacts network~\cite{liljeros2001}. Some  studies also discuss the behavior of the network under random removal of its links or nodes~\cite{allard2009}.

In recent work, random link removal on monopartite graphs was extended to \textit{biased percolation}, in which links are removed according to the degree of their neighboring nodes~\cite{moreira}. More specifically, to each link is attached a weight
\begin{equation}
w_{kq} = (k q)^{-\alpha} \label{wij}
\end{equation}
where $k$ and $q$ are the neighboring node degrees and $\alpha$ is the bias exponent which we take to be positive (and $< 1$) for our present purposes~\cite{noot}. During the depreciation process,
the probability to {\it retain} a link is proportional to the weight of the link. Since $\alpha >0$, links between hubs are more prone to failure. Therefore, the percolation process is called \textit{centrally biased}. Such a process is inherent to many social networks, where friendships between people with many acquaintances are expected to be weaker and last less long than  friendships involving people with few connections \cite{Indekeu}.
%  Also in technological networks, like the air transportation network, links between hubs are more prone to failure, since the excess of traffic makes the highly loaded links more vulnerable.
In previous articles, Refs.~\cite{moreira} and~\cite{hooyberghs2009}, we showed that the critical behavior of such a biased percolation on a scale-free network can be mapped onto a random removal process on another scale-free network. More precisely, biased percolation with bias exponent $\alpha > 0$ on a network with topological exponent $\gamma$ belongs to the same universality class as random percolation ($\alpha = 0$) on a network with topological exponent $\overline{\gamma}$, given by
\begin{equation}
 \label{gammaac} \overline{\gamma} = \frac{\gamma -\alpha}{1-\alpha}.
\end{equation}
Therefore, the network is, in the macroscopic limit, robust against biased removal of links as long as $2<\overline{\gamma}<3$. If $\overline{\gamma} > 3$, the network is fragile for the central bias process. Note that $\overline{\gamma} > \gamma$.

The paper at hand discusses biased percolation on bipartite scale-free networks. In Sect.~2 we introduce bipartite scale-free networks and present a basic description of a link removal process on those networks. The third section introduces the theory of generating functions, which we extend so as to describe bipartite networks. The percolation threshold and the scaling of the critical point are extracted from the theory. In Sect.~4, the equivalence between percolation and a limit of the Potts model is elaborated. Using this equivalence, we construct a finite-size scaling theory which enables the calculation of the critical exponents for the percolation transition. In Sect.~5 we present our conclusions.

\section{THE MODEL}
We start from a bipartite network with $N$ nodes, divided in $N_A$ nodes of type $A$ and $N_B$ nodes of type $B$, each type with its own degree distribution $P^A(k)$ and $P^B(k)$. Both degree distributions are assumed to follow a decreasing power law and thus are scale free. More precisely,
\begin{equation}
 P^i(k) = C_i k^{-\gamma_i} \label{jkl}
\end{equation}
for values of $k$ between the minimal and maximal degrees, $m_i$ and $K_i$, respectively, for $i=A,B$. In Eq.~(\ref{jkl}), $C_i$ is a normalization constant and the exponent $\gamma_i$ is assumed to be larger than two to ensure a finite mean degree.
Links exist only between nodes of unlike type. Moreover, we assume that no degree correlations or mixing patterns between the nodes, as for instance  assortative mixing, occur. The probability $P^A_n(k)$ that a randomly chosen edge emerging from a node of type $A$ leads to a type-$B$ node with degree $k$ thus only depends on the degree distribution $P^B(k)$. More precisely,
\begin{equation} \label{winterweer}
 P^A_n(k) = \frac{kP^B(k)}{\avk_B},
\end{equation}
where $\langle\cdot \rangle_i$ denotes the average over the nodes of type $i$, obtained by using the degree distribution $P^i(k)$. Note that the total number of links attached to the nodes of type $A$ equals the total number of links attached to type-$B$ nodes, i.e.~$\avk_A N_A = \avk_B N_B$.

In our percolation process, a fraction $f$ of the links is removed in a single sweep; we call this the {\it simultaneous} process. An edge between nodes with degrees $k$ and $q$ is retained with probability
\begin{equation}\label{sporza}
 \rho_{kq} = f\frac{w_{kq}}{\langle w \rangle_e},
\end{equation}
and is removed with probability $1-\rho_{kq}$. The weight $w_{kq}$ is defined in Eq.~(\ref{wij}) and $\langle w \rangle_e$ denotes the average weight of an edge,
\begin{equation}
 \langle w \rangle_e = \frac{\average{1-\alpha}_A\average{1-\alpha}_B}{\avk_A\avk_B}.
\end{equation}
Note that the random link removal process is recovered if $\alpha=0$. The positive bias exponent $\alpha$ should be smaller than one in order for the depreciated network to be scale free.  Moreover, the  depreciation process is only well-defined if $\rho_{kq} <1$ for all possible values of $k$ and $q$. Therefore, our percolation can only be used correctly for values of $f$ for which~\cite{alternatief}
\begin{equation}
 f < f_u =  \langle w \rangle_e (m_Am_B)^\alpha.
\end{equation}

We now introduce the marginal distribution $\rho^i_k(f)$ as the mean probability that an edge connected to a node of type $i$ with degree $k$ is present when a fraction $f$ of links is reincluded in the network. One finds
\begin{equation}
 \rho^A_k(f) = \frac{f\avk_A}{\average{1-\alpha}_A}k^{-\alpha}. \label{labello}
\end{equation}
Note that the marginal distribution $\rho^A_k$ does not depend on the distribution of the nodes of type $B$ and vice versa. Using the marginal distribution, the degree distribution $\overline{P}^i(\overline{k})$ and nearest-neighbor degree distribution $\overline{P}^i_n(\overline{k})$ of nodes of type $i$ in the diluted network can be deduced. One arrives at
\begin{subequations} \label{zonneschijn}
\begin{eqnarray}
\overline{P}^i(\overline{k})&=& \sum_{ k =\overline{k}}^{K_i} P^i(k)\left(\begin{array}{c}
                                                                     ×k\\\overline{k}
                                                                    \end{array}\right) (\rho^i_k)^{\overline{k}}
        (1-\rho^i_k)^{k-\overline{k}},\\
\overline{P}_n^i(\overline{k})&=& \frac{k\overline{P}^j(\overline{k})}{f\avk_j}.
\end{eqnarray}
\end{subequations}
The importance of these expressions will become clear when introducing the generating functions in the next section. Moreover, it can be shown that
\begin{equation}
 f\rho_{kq} = \rho^A_k\rho^B_q,
\end{equation}
and thus, according to the arguments in Ref.~\cite{hooyberghs2009}, the diluted network is still uncorrelated.

\section{GENERATING FUNCTIONS}
Percolation is often studied using the generating functions approach. By this method, self-consistent equations for the size of the giant cluster can be obtained easily. Moreover, the method is exact if the diluted network is uncorrelated and if loops in the finite clusters can be ignored, which is justified for scale-free networks~\cite{bianconi2005}. We first briefly introduce the general scheme, closely following the approach of Newman~\cite{newman2}. Then, the generating functions method is exploited for the case of biased percolation on bipartite networks.
\subsection{Self-consistent equations}
Generating functions are used in a plethora of problems concerning series. The generating function of a sequence is the power series which has as coefficients the elements of the sequence~\cite{wilf}. In percolation problems, generating functions that generate the probability distributions characterizing the network are widely used~\cite{newman2,essam}. The most important functions are those that generate the degree distribution and those that generate the nearest-neighbor distribution. To study the percolation problem, both generating functions will be defined in the diluted network, for both vertex types.

%
% The generating function for the degree distribution $P^i(k)$ is defined as
% \begin{equation}
%  G_0^i(h) = \sum_{k=m_i}^{K_i} P^i(k) e^{-hk},\label{G0}
% \end{equation}
% while the distribution of the residual edges is generated by
% \begin{equation}
% G_1^i(h) = \sum_{k=m_i}^{K_i} P^i_n(k)e^{-h(k-1)}, \label{G10}
% \end{equation}
% where $i = A,B$.

%\subsection{Self-consistent equations}\label{selfconsistent}
% To study the percolation problem, we now define the equivalents of $G^i_0(h)$ and $G^i_1(h)$ for the
% network \textit{after dilution} as $F^i_0(h)$ and $F^i_1(h)$:
% \begin{eqnarray}\label{genererendefunctiesrule}
% F_0^{i}(h)&=&\sum_{\overline{k}=1}^{K_i} \overline{P}^{i}(\overline{k}) e^{-h\overline{k}},\\
% F_1^{i}(h)&=&\sum_{\overline{k}=1}^{K_i}
% \overline{P}_n^{i}(\overline{k})e^{-h(\overline{k}-1)}.
% \end{eqnarray}

The generating function for the degree distribution $\overline{P}^i(k)$ is defined as
% \begin{subequations}
\begin{equation}\label{genererendefunctiesrule}
F_0^{i}(h)=\sum_{\overline{k}=1}^{K_i} \overline{P}^{i}(\overline{k}) e^{-h\overline{k}},
\end{equation}
while the distribution of the residual edges in the diluted network is generated by
\begin{equation}
F_1^{i}(h)=\sum_{\overline{k}=1}^{K_i}
\overline{P}_n^{i}(\overline{k})e^{-h(\overline{k}-1)},
\end{equation}
% \end{subequations}
where $i = A,B$.
Substituting Eqs.~(\ref{zonneschijn}) and~(\ref{winterweer}), we obtain
\begin{subequations}
\begin{eqnarray}
F_0^{i}(h) &=& \sum_{k=m_i}^{K_i} P^{i}(k) (1- \rho_k^i + e^{-h}\rho_k^i)^k\label{ikke},\\
F_1^{i}(h) &=& \sum_{k=m_i}^{K_i}  P_n^{i}(k) (1- \rho_k^i + e^{-h}\rho_k^i)^{k-1}\label{ikkeb}
\label{F1},
\end{eqnarray}
\end{subequations}
with $i=A,B$. For our interest, the most relevant generating functions for the percolation problem are the ones associated with the probability distribution of the size of the finite clusters, since these quantities can be related to the size of the giant cluster.
Let $H^i_0$ generate the probability that a randomly chosen node of type $i$ belongs to a cluster of a given finite size. Furthermore, we introduce $H^i_1$ as the generating function for the
probability that upon following a randomly chosen edge emerging from a node $i$ towards the endnode of type $j$,
a cluster of given (finite) size is reached. If the finite clusters can be
treated as trees and the diluted network is uncorrelated, these generating functions satisfy coupled self-consistency equations, analogous to those derived for monopartite graphs in Ref.~\cite{newman2}. For bipartite graphs, we obtain
\begin{subequations}\label{zon}
\begin{eqnarray}
H_1^{A}(h) &=& e^{-h}F_1^{B}[H_1^{B}(h)],\label{H0}\\
H_1^{B}(h) &=& e^{-h}F_1^{A}[H_1^{A}(h)],\label{H0bis}\\
H_0^{A}(h) &=& e^{-h}F_0^{A}[H_1^{A}(h)],\\
H_0^{B}(h)&=& e^{-h}F_0^{B}[H_1^{B}(h)].\label{H1}
\end{eqnarray}
\end{subequations}
Here the function $F_1^{i}[H_1^i(h)]$ denotes the function
$F_1^{i}$ wherein $e^{-h}$ is replaced by $H_1^{i}(h)$ with
$i=A,B$. The percolation threshold can now be derived with the aid of these generating
functions and self-consistent relations.

\subsection{Percolation threshold}
The percolation threshold is most easily studied by introducing the average cluster size in the diluted network, i.e., the average fraction of nodes of type $i$ in a cluster, denoted by $\mathcal{S}^i$. Using the properties of generating functions, $\mathcal{S}^i$ can be related to $H^i_0$~\cite{wilf}:
\begin{equation}
 \mathcal{S}^i=-\dot{H}_0^{i}(0),
\end{equation}
where the dot represents differentiation with respect to $h$.
An expression for the average cluster size $\mathcal{S}$, i.e., the average fraction of nodes of type A and B in a finite cluster, is then easily found:
\begin{equation}
 \mathcal{S} = -\frac{1}{2}\sum_{i=A,B}\dot{H}_0^{i}(0)\label{timothy}.
\end{equation}
The average cluster size $\mathcal{S}$ in the diluted network can be further worked out by differentiating
Eqs.~(\ref{ikke}) and (\ref{ikkeb}) with respect to $h$:
\begin{equation}
\mathcal{S}  =  1 + \frac{f}{2}\frac{\avk_A(1-\dot{F}_1^B(0))+\avk_B(1-\dot{F}_1^A(0))}{1
-\dot{F}_1^A(0)\dot{F}^B_1(0)}. \label{avcl}
\end{equation}
Hence the average cluster size diverges when
\begin{equation}
 1 = \dot{F}_1^A(0)\dot{F}^B_1(0). \label{crit}
\end{equation}
The percolation criterion (\ref{crit}) is the extension for bipartite graphs of the Molloy-Reed criterion, which provides a condition for the existence of a giant cluster in a network~\cite{cohen,molloy}. A similar expression for the percolation threshold was already found in studies concerning %virus spreading on sexual contacts networks~\cite{newman} and
 percolation on general multipartite networks~\cite{allard2009}.

\begin{figure*}[t!]
\caption{Overview of the different universality regimes as a function of the scale free exponents $\overline{\gamma}_A$ and $\overline{\gamma}_B$. The red line indicates the division between a fragile network (red dotted regimes) and a robust network (blank regimes). \label{fasediagram}}

\epsfig{figure=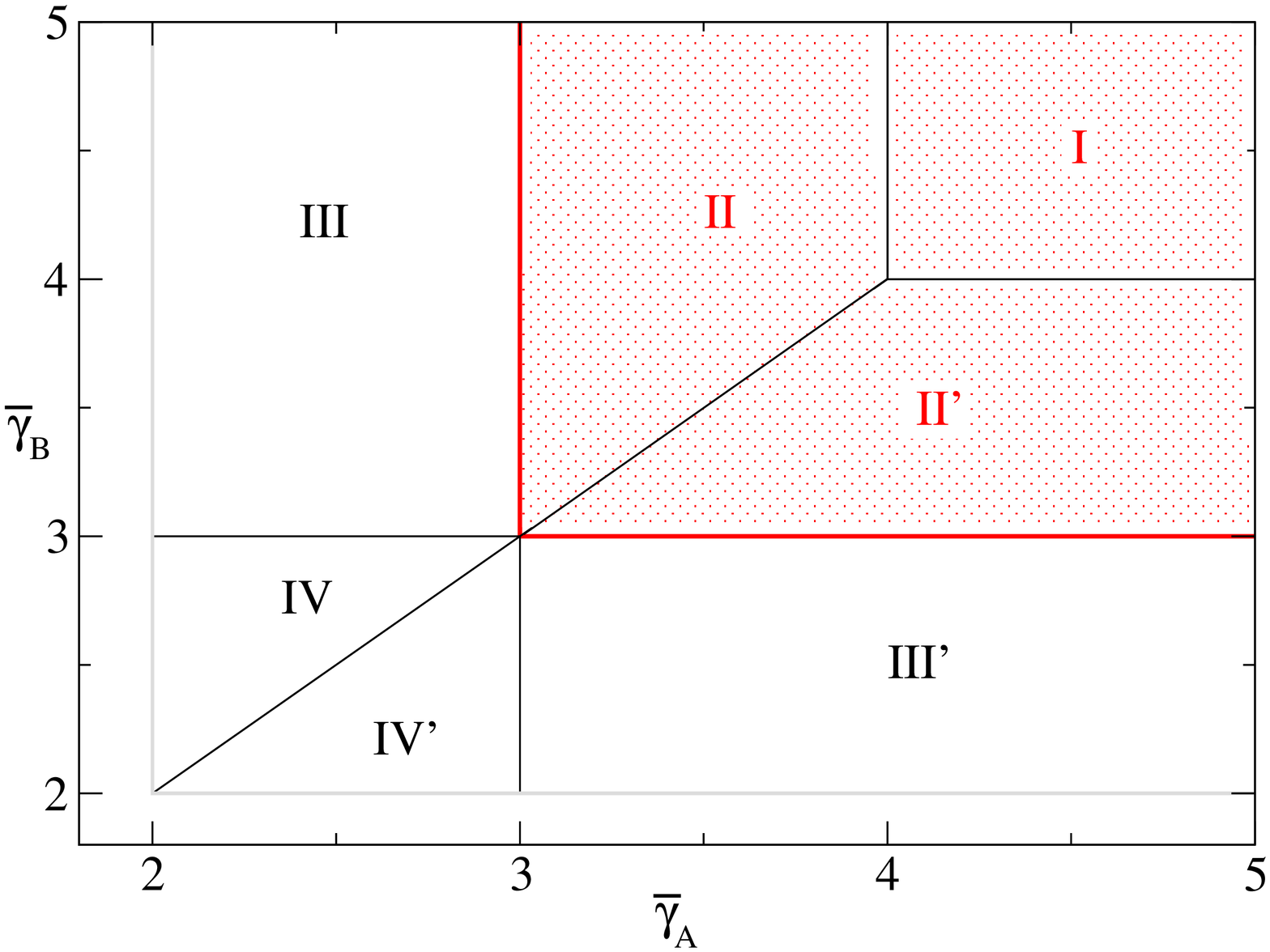,angle=0, width = .75\textwidth}
\end{figure*}

Using Eqs.~(\ref{labello}) and~(\ref{ikkeb}) an explicit criterion for the critical fraction $f_c$ can be found,
\begin{equation}
  f_c = \sqrt{f_c^A f_c^B}, \label{onscrit}
\end{equation}
where $f_c^i$ denotes the critical fraction of a monopartite graph consisting only of nodes of type $i$ with degree distribution $P^i(k)$,
\begin{equation}\label{formule1}
 f_c^i = \frac{\average{1-\alpha}_i^2}{\avk_i\left(\average{2-2\alpha}_i - \average{1-2\alpha}_i\right)}.
\end{equation}
Note that Eq.~(\ref{onscrit}) reduces to the criterion for monopartite graphs if $\gamma_A=\gamma_B$, as it should be. Moreover, the critical fraction of a bipartite graph can easily be found if the critical fraction of the monopartite graphs consisting of only nodes of type  $A$ and $B$ is known. As a consequence, also the resilience against biased failures of the bipartite network in the macroscopic limit is known completely if the behavior of the different subgraphs is known. If one of the two subgraphs is robust against biased percolation, also the bipartite network will be robust. The robustness criterion is therefore:
\begin{equation}
\min\left(\overline{\gamma}_A,\overline{\gamma}_B\right) < 3,
\end{equation}
where $\overline{\gamma}_i$ is defined as in Eq.~(\ref{gammaac}). This is illustrated in Fig.~\ref{fasediagram}, where the robust regimes are indicated by the blank (undotted) regions. Only if both $\overline{\gamma}_A$ and $\overline{\gamma}_B$ are larger than 3, the network is fragile, indicated by the regimes dotted in red on Fig.~1.

For the robust regimes, we can quite easily determine the scaling relation  of $f_c$ as a function of the network size, $f_c \propto N^\varsigma$, by explicitly evaluating the expectation values in Eq.~(\ref{formule1}). Replacing the sums over the degree distibution by integrals, it can be seen that the moments $\avk_i$ and $\average{1-\alpha}_i$ never diverge if central bias is applied to a network with $\gamma_i>2$. The moments $\average{2-2\alpha}_i$ and $\average{1-2\alpha}_i$, on the other hand, may diverge, but the former will always grow faster than the latter.  The behavior of the critical fraction as a function of the maximal degrees, $K_i$ for $i=A,B$, thus stems from the first term, which for $2<\gamma_i<3$ diverges as $K_i^{3-2\alpha-\gamma_i}$. Since $K_i \propto N_i^{1/(\gamma_i - 1)}$~\cite{cohen}, the scaling of the critical point as a function of the network size can be determined. If both $\overline{\gamma}_A$ and $\overline{\gamma}_B$ are smaller than 3 (regimes IV and IV' on Fig.~\ref{fasediagram}), $\varsigma$ is given by
\begin{equation}
\varsigma = \frac{1}{2}\left(\frac{3-\overline{\gamma}_A}{1-\overline{\gamma}_A} + \frac{3-\overline{\gamma}_B}{1-\overline{\gamma}_B}\right).
\end{equation}
Note that this reduces to the result for monopartite graphs, determined in Ref.~\cite{moreira}, if $\gamma_A = \gamma_B$.
If $2<\gamma_A <3 < \gamma_B$ (regime III on Fig.~1), the exponent $\varsigma$ is the same as the exponent for a robust monopartite graph of type~A:
\begin{equation}\label{hierboven}
\varsigma = \frac{3-\overline{\gamma}_A}{1-\overline{\gamma}_A}.
\end{equation}
Note that the exponent in the regime III' ($2<\gamma_B <3 < \gamma_A$) can be found easily by interchanging $\gamma_A$ and $\gamma_B$ in Eq.~(\ref{hierboven}).

In sum, we have now provided an extension of the Molloy-Reed criterion for the critical threshold of biased percolation on bipartite scale-free networks. Moreover, the scaling of the critical fraction as a function of the network size was determined for robust graphs. Our results show that the critical fraction of a bipartite network is governed by the critical percolation behavior of the scale-free graphs consisting of only one type of nodes.

\section{CRITICAL BEHAVIOR}
In the following section, we introduce and calculate the critical exponents of the percolation transition.
The first part briefly introduces the Fortuin-Kasteleyn construction, which provides a link between percolation and a limit of the Potts model. Exploiting this link with the Potts model, we can define critical exponents for the percolation problem. In the second part of the section, a finite-size scaling theory is constructed, in order to calculate the critical exponents in the third part.

\subsection{Fortuin-Kasteleyn construction}\label{kastfort}
There exists an exact equivalence between random edge percolation and the $q\rightarrow 1$ limit of the $q$-state Potts model, originally worked out by Fortuin and Kasteleyn in Ref.~\cite{fortuin}.  Moreover, their proof can easily be generalized to incorporate edge-dependent coupling constants in the Potts model and edge-dependent removal in the percolation model, respectively. The Fortuin-Kasteleyn construction states that the free energy of the $q\rightarrow 1$ limit of the $q$-state Potts model is the same as the ``free energy'' of the percolation problem~\cite{wu}. The latter is defined as the generating function of the cluster-size distribution $n_s$,
\begin{equation}
\mathcal{F}(f,h)=\left\langle \sum_s n_s e^{-hs}\right\rangle.
\end{equation}
Here the average is performed over all networks in which the
probability to retain the edge between nodes with degrees $k$ and $q$ is
$\rho_{kq}$. We can immediately identify the probability
$\mathcal{P}_{\infty}$ for a node of any type to be in the infinite cluster
and the average cluster size $\mathcal{S}$, as a function of the fraction of removed links:
\begin{subequations}
\begin{eqnarray}
\mathcal{P}_{\infty}(f)&=&1+\left.\frac{\partial
\mathcal{F}}{\partial
h}\right|_{h=0},\\
\mathcal{S}(f)& =&\left.\frac{\partial^2
\mathcal{F}}{\partial h^2}\right|_{h=0}.
 \end{eqnarray}
\end{subequations}
% Note also that $\mathcal{F}(\epsilon,0)$ gives the total amount of
% finite clusters.
Since we are interested in the behavior close to criticality, we introduce the parameter
\begin{equation}
\epsilon = f - f_c.
\end{equation}
The usual critical exponents $\alpha$, $\gamma_p$, $\beta$ and $\delta$ can now be defined for
the percolation problem using the analogy with the Potts model:
\begin{subequations}
\begin{eqnarray}
\mathcal{F}(\epsilon,0)&\sim&
\epsilon^{2-\alpha},\\
\mathcal{P}_{\infty}(\epsilon)&\sim&
\epsilon^{-\gamma_p},\\
\mathcal{S}(\epsilon)&\sim&
\epsilon^{\beta},\\
\left.\frac{\partial \mathcal{F}}{\partial
h}\right|_{\epsilon=0} &\sim
& h^{1/\delta}-1.
\end{eqnarray}
\end{subequations}
Relations among these exponents can be found using a scaling theory.

\subsection{Scaling Theory}\label{critexponents}
In the following section, we introduce finite-size scaling in
order to find critical exponents near the percolation transition.
In order to solve the scaling relation, we use a Landau-like
theory which we derive from the exact self-consistent relations, Eqs.~(\ref{zon}). We
closely follow the approach presented in Ref.~\cite{hooyberghs2009}, which is based on Refs.~\cite{hong,botet1,botet2}.
% However, we will find that the forms of the Landau-like theories
% of Refs.~\cite{hong,igloi,goltsev2} were too limited for studying
% the percolation transition in case the distribution function has a
% very fat tail, that is when $2<\overline{\gamma}<3$.

Our scaling theory consists of two basic scaling relations. The free energy $\mathcal{F}$ of
a large but finite network with $N$ nodes close to criticality can
be written in the general form~\cite{stauffer}:
\begin{equation}
 \label{freeenergy}
\mathcal{F}(\epsilon,h)=N^{-1}\mathbb{F}\left(\epsilon
N^{1/\nu_{\epsilon}},hN^{1/\nu_{h}}\right),
\end{equation}
where $\mathbb{F}$ is a well-behaved function. Close to the critical point, the free energy then scales as
\begin{subequations}\label{denieuwe}
 \begin{eqnarray}
  \mathcal{F}(\epsilon,0) &\propto&\epsilon^{\nu_\epsilon},\\
  \mathcal{F}(0,h) &\propto&h^{\nu_h}.
 \end{eqnarray}
\end{subequations}
As a second ansatz, the scaling of the cluster size distribution $n_s$ can in the thermodynamic limit be
written as
\begin{equation}
 \label{clusterdistribution}
n_s(\epsilon)=s^{-\tau}\mathbb{G}(\epsilon s^{\sigma}).
 \end{equation}
Using the scaling forms of Eqs.~(\ref{freeenergy})
and~(\ref{clusterdistribution}), standard techniques provide us
with exponent relations by which all critical exponents can be
related to $\nu_{h}$ and $\nu_{\epsilon}$. One arrives
at~\cite{stauffer}:
\begin{subequations}\label{exponents}
\begin{eqnarray}
\beta &=&\nu_{\epsilon}(1-\nu_{h}^{-1}),\\
\gamma_p&=&\nu_{\epsilon}(2\nu_{h}^{-1}-1),\\
\alpha &=&2-2\beta-\gamma_p,\\
\sigma &=&(\beta+\gamma_p)^{-1},\\
\tau &=&2+\beta(\beta+\gamma_p)^{-1},\\
\delta &=&(\beta+\gamma_p)/\beta.\label{exponentsend}
\end{eqnarray}
\end{subequations}
The problem we are left with is to find the exponents
$\nu_{h}$ and $\nu_{\epsilon}$ for percolation on bipartite scale-free
networks.

\subsection{Critical exponents}
Using the self-consistent equations, the exponents $\nu_h$ and $\nu_\epsilon$ can be determined in an exact way. We introduce the parameters
\begin{equation}
\psi_i(\epsilon,h)=1-H_1^i(\epsilon,h),
\end{equation}
where $i=A,B$.
As we are merely interested in the behavior near the transition
where $\epsilon\ll 1$, $h\ll 1$ and $\psi_i\ll 1$, we can expand
Eqs.~(\ref{H0}) and~(\ref{H0bis}).
\subsubsection{Fragile networks}
We first discuss all regimes in which both $\overline{\gamma}_A$ and $\overline{\gamma}_B$ are larger than three, i.e., the regimes in which the network is fragile and has a finite $f_c$. An expansion of the self-consistent equations, Eqs. (\ref{zon}), yields
\begin{subequations}\label{eos}\label{de36}
\begin{eqnarray}
h-\psi_B &=&-c_1^A(f_c+\epsilon) \psi_A
+c_2^A\psi_A^2+\ldots  + c_s^A(\psi_A)^{\overline{\gamma}_A-2}+\ldots,\\
h-\psi_A &=&-c_1^B(f_c+\epsilon) \psi_B
+c_2^B\psi_B^2+\ldots + c_s^B(\psi_B)^{\overline{\gamma}_B-2}+\ldots,
\end{eqnarray}
\end{subequations}
wherein all $c$-constants are positive and according to the
percolation criterion $f_c^2c_1^Ac_1^B=1$. The existence of the correction terms, $c_s^i(\psi_A^i)^{\overline{\gamma}_i-2}$, can be checked numerically. Eqs.~(\ref{de36}) are
derivable by minimization of the free energy
\begin{eqnarray}\label{tristan}
\frac{\mathcal{F}}{f}&=&\psi_A\psi_B+\sum_{i=a,b}\left(-h\psi_i-\frac{c_1^i(f_c+\epsilon)\psi_i^2}{2}
+\frac{c_2^i\psi_i^3}{3}+\frac{c_s^i(f_c+\epsilon)^{\overline{\gamma}_i-2}\psi_i^{\overline{\gamma}_i-1}}{\overline{\gamma}_i-1}\right) + \ldots,
 \end{eqnarray}
with respect to the parameters $\psi_i$.
Using the equations of state, Eqs.~(\ref{eos}), we may write at the saddle point
\begin{eqnarray}
\frac{\mathcal{F}}{f_c}&=&\sum_{i=a,b}\left(-\frac{h\psi_i}{2}
-\frac{c_2^i\psi_i^3}{6}+\frac{c_s^i(3-\overline{\gamma}_i)}{2(\overline{\gamma}_i-1)}(f_c+\epsilon)^{\overline{\gamma}_i-2}\psi_i^{\overline{\gamma}_i-1}\right). \label{dewindwaait}
 \end{eqnarray}
 If $\overline{\gamma}_A$ and  $\overline{\gamma}_B$ are both larger than 4, i.e., in regime I in Fig.~\ref{fasediagram}, the equations of state reduce in lowest order to
\begin{subequations}
\begin{eqnarray}
 h-\psi_B &=&-c_1^A(f_c+\epsilon) \psi_A
+c_2^A\psi_A^2,\\
 h-\psi_A &=&-c_1^B(f_c+\epsilon) \psi_B +c_2^B\psi_B^2.
\end{eqnarray}
\end{subequations}
Solving for $\psi_A$ and $\psi_B$, we obtain
\begin{subequations}
\begin{eqnarray}\label{robert}
\psi_A|_{h=0}\sim \psi_B|_{h=0}&\sim&\epsilon,\\
\psi_A|_{\epsilon=0}\sim \psi_B|_{\epsilon=0}&\sim& h^{1/2}.
\end{eqnarray}
\end{subequations}
Substitution into Eq.~(\ref{dewindwaait}) yields
\begin{subequations}
\begin{eqnarray}
\mathcal{F}(\epsilon,0)&\sim& \epsilon^{3},\\
\mathcal{F}(0,h)&\sim& h^{3/2},
\end{eqnarray}\end{subequations}
thus, according to Eqs.~(\ref{denieuwe}), $\nu_\epsilon = 3$ and $\nu_h = 3/2$. All other exponents can now be calculated using Eqs.~(\ref{exponents}). We list the result in the first column of Table \ref{tabel}.

Next, we focus on the regime in which one of the topological exponents is larger than four, while the other exponent has a value between three and four. Without loss of generality, we take in the remainder of the text $\overline{\gamma}_A$ as the smallest of the two exponents. The self-consistent equations now reduce to
\begin{subequations}
\begin{eqnarray}
 h-\psi_B &=&-c_1^A(f_c+\epsilon) \psi_A
+c_s^A\psi_A^{\overline{\gamma}_A-2},\\
 h-\psi_A &=&-c_1^B(f_c+\epsilon) \psi_B +c_2^B\psi_B^2,
\end{eqnarray}
\end{subequations}
from which we obtain
\begin{subequations} \label{dsf}
\begin{eqnarray}
 \nu_\epsilon &=& \frac{1}{\overline{\gamma}_A-3},\\
 \nu_h &=& \frac{1}{\overline{\gamma}_A-2}.
 \end{eqnarray}
\end{subequations}
All other critical exponents are given in the second column of Table \ref{tabel}.

Finally, we discuss the exponents in the case $3<\overline{\gamma}_A<\overline{\gamma}_B<4$.
Although the self-consistent equations are slightly different from those in the previous paragraph, we can verify that Eqs.~(\ref{dsf}) are still valid. The exponents in this regime are thus the same as in the regime discussed in the previous paragraph. Therefore, we define region II as the region in which $3< \overline{\gamma}_A<4$ and $ \overline{\gamma}_A < \overline{\gamma}_B$, as is illustrated in Fig.~\ref{fasediagram}. The exponents in that regime can be found in the second column of Table~\ref{tabel}.

\subsubsection{Robust networks}
We still have to discuss the exponents for robust bipartite networks. We first focus on regime IV in Fig.~1 where both $\overline{\gamma}_A$ and $\overline{\gamma}_B$ are smaller than three. Since $f_c = 0$, the self-consistent equations close to the critical point reduce to
\begin{subequations}
 \label{eoshans}
\begin{eqnarray}
h-\psi_B &=&-c_s^A(\epsilon\psi_A)^{\overline{\gamma}_A-2},\\
h-\psi_A &=&-c_s^B(\epsilon\psi_B)^{\overline{\gamma}_B-2},
 \end{eqnarray}
\end{subequations}
where $c_s^A>0$ and $c_s^B>0$. These equations can be derived by minimization of
\begin{equation}
 \frac{\mathcal{F}}{\epsilon}=\psi_A\psi_B+\sum_{i=A,B}-h\psi_i
 -\frac{c_s^i\epsilon^{\overline{\gamma}_i-2}\psi_i^{\overline{\gamma}_i-1}}{\overline{\gamma}_i-1},
\end{equation}
with respect to the order parameters. At the saddle point, we may write:
\begin{equation}
 \frac{\mathcal{F}}{\epsilon}=\sum_{i=a,b}-\frac{h\psi_i}{2}
-\frac{c_s^i(3-\overline{\gamma}_i)}{2(\overline{\gamma}_i-1)}\epsilon^{\overline{\gamma}_i-2}\psi_i^{\overline{\gamma}_i-1}. \label{kael}
\end{equation}
By solving Eqs.~(\ref{eoshans}) for $\psi_i$, one obtains
\begin{subequations}
\begin{eqnarray}
\psi_A|_{h=0}&\sim& \epsilon^{\frac{(\overline{\gamma}_A-1)(\overline{\gamma}_B-2)}{1-(\overline{\gamma}_A-2)(\overline{\gamma}_B-2)}},\\
\psi_B|_{h=0}&\sim&\epsilon^{\frac{(\overline{\gamma}_B-1)(\overline{\gamma}_A-2)}{1-(\overline{\gamma}_A-2)(\overline{\gamma}_B-2)}},\\
\psi_A|_{\epsilon\rightarrow0}&\sim& |h|^{\frac{1}{\overline{\gamma}_A-2}}/\epsilon,\\
\psi_B|_{\epsilon\rightarrow0}&\sim& |h|^{\frac{1}{\overline{\gamma}_B-2}}/\epsilon.
\end{eqnarray}
\end{subequations}
Substitution into Eq.~(\ref{kael}) yields
\begin{subequations}
\begin{eqnarray}
 \nu_\epsilon &=& \frac{(\overline{\gamma}_A-1)(\overline{\gamma}_B-1)}{1-(\overline{\gamma}_A-2)(\overline{\gamma}_B-2)},\\
 \nu_h &=& \frac{\overline{\gamma}_B-1}{\overline{\gamma}_B-2}.
\end{eqnarray}
\end{subequations}
The other exponents are listed in the last column of Table 1.

Finally, we focus on the regime in which $\overline{\gamma}_A <3$ and $\overline{\gamma}_B > 3$, i.e. regime III on Fig.~1. Although the expansion of the self-consistent relations, Eqs. (\ref{zon}), depends on the actual value of $\overline{\gamma}_B$, the scaling of the order parameters does not.
After some calculations, we obtain
% \begin{eqnarray}
%  \psi_A &\sim& \epsilon^{\frac{\overline{\gamma}_A-1}{3 - \overline{\gamma}_A}},\\
%  \psi_B &\sim& \epsilon^{\frac{2(\overline{\gamma}_A-2)}{3 - \overline{\gamma}_A}},\\
%  \psi_A &\sim& (h/\epsilon)^\frac{1}{\overline{\gamma}_A-2}/\epsilon,\\
%  \psi_B &\sim& h/\epsilon.
% \end{eqnarray}
% Some calculations yield
\begin{subequations}
\begin{eqnarray}
 \nu_\epsilon &=& \frac{2(\overline{\gamma}_A - 1)}{3 - \overline{\gamma}_A},\\
 \nu_h &=& 2.
\end{eqnarray}
\end{subequations}
We list the other exponents in the third column of Table 1.
\subsubsection{Discussion}
To summarize, we have calculated the critical exponents of the percolation transition for all physically relevant regimes. Note that all exponents in the regimes I, II and IV reduce to the correct expressions for monopartite graphs if $\overline{\gamma}_A = \overline{\gamma}_B$~\cite{hooyberghs2009}. Note also that, as expected, the usual mean-field results for percolation are recovered only in regime I~\cite{essam,stauffer}. In all other regimes, we find non-universal exponents which depend on the constants $\overline{\gamma}_i$.

The only dependence on the bias exponent $\alpha$ arises through the exponents $\overline{\gamma}_i$ as defined in Eq.~(\ref{gammaac}). Biased percolation with bias exponent $\alpha$ on a bipartite scale-free network with topological exponents $\gamma_A$ and $\gamma_B$ thus has the same critical behavior as random percolation ($\alpha = 0$) on a network with topological exponents $\overline{\gamma}_A$ and $\overline{\gamma}_B$. We thus conclude that the results for bipartite graphs are a generalization of those for monopartite graphs.

\begin{table*}[t!]
\caption{Critical exponents in the different regimes for which $\overline{\gamma}_A <\overline{\gamma}_B$. Note that the exponents for the primed regimes in Fig.~\ref{fasediagram} can be obtained by interchanging $\overline{\gamma}_A$ and $\overline{\gamma}_B$ in the expressions for the regimes without primes. \label{table1}}
\ra{1.7}
\newcommand{\cc}[1]{\multicolumn{1}{c}{#1}}
\renewcommand{\tabcolsep}{1.pc}
\begin{tabular}{@{}lcccc}
\hline
& \cc{Regime I} & \cc{Regime II} & \cc{Regime III} & \cc{Regime IV}\\
\hline
$\varsigma$
&Fragile&Fragile&$\frac{3-\overline{\gamma}_A}{1-\overline{\gamma}_A}$ &$\frac{1}{2}\left(\frac{3-\overline{\gamma}_A}{1-\overline{\gamma}_A} + \frac{3-\overline{\gamma}_B}{1-\overline{\gamma}_B}\right)$\\
$\beta$
& $1$ & $\frac{1}{\overline{\gamma}_A-3}$ & $\frac{\overline{\gamma}_A-1}{3-\overline{\gamma}_A}$ & $\frac{\overline{\gamma}_A-1}{1-(\overline{\gamma}_A-2)(\overline{\gamma}_B-2)}$\\

$\tau$ & $5/2$& $\frac{2\overline{\gamma}_A-3}{\overline{\gamma}_A-2}$& $3$& $\frac{2\overline{\gamma}_B-3}{\overline{\gamma}_B-2}$\\

$\sigma$ & $1/2$& $\frac{\overline{\gamma}_A-3}{\overline{\gamma}_A-2}$& $\frac{3-\overline{\gamma}_A}{\overline{\gamma}_A-1}$
& $\frac{1-(\overline{\gamma}_A-2)(\overline{\gamma}_B-2)}{(\overline{\gamma}_A-1)(\overline{\gamma}_B-2)}$\\

$\alpha$& $-1$& $-\frac{5-\overline{\gamma}_A}{\overline{\gamma}_A-3}$& $-\frac{4(\overline{\gamma}_A-2)}{3-\overline{\gamma}_A}$
&$-\frac{7+3\overline{\gamma}_A\overline{\gamma}_B-5(\overline{\gamma}_A+\overline{\gamma}_B)}{1-(\overline{\gamma}_A-2)(\overline{\gamma}_B-2)}$\\

$\overline{\gamma}_p$ & $1$& $1$& $0$
& $-\frac{(3-\overline{\gamma}_B)(\overline{\gamma}_A-1)}{1-(\overline{\gamma}_A-2)(\overline{\gamma}_B-2)}$\\

$\delta$& $2$& $\overline{\gamma}_A-2$& $1$& $\overline{\gamma}_B-2$\\
\hline

\end{tabular}
\label{tabel} % is used to refer this table in the text
\end{table*}

\section{CONCLUSION}
As an extension of previous work on scale-free graphs with a single type of nodes~\cite{moreira,hooyberghs2009}, this article studies the biased removal of links in a bipartite network. A bipartite network consists of two types of nodes, with links only running between nodes of unlike type. We assume that the degrees of both nodes are distributed according to a decreasing power law, without any correlations or mixing patterns between the two types of nodes. This model can be used to describe real-life social networks, such as for instance the network of hetero-sexual contacts~\cite{liljeros2001}. In our percolation process, we reinstall a fraction $f$ of the links in a single sweep. Moreover, the process is biased in the sense that the link removal probability depends on the degrees of the nodes they connect. We attach a weight $(kq)^{-\alpha}$ to a link between nodes with degrees $k$ and $q$ and retain links with a probability proportional to their weights. In the study at hand, the bias exponent $\alpha$ is a positive number smaller than one. Therefore, links between hubs are more prone to failure, thus the process is centrally biased.

The most important conclusion of the article is that almost all results for the bipartite graphs are extensions of the results for monopartite graphs. The robustness or fragility of a bipartite network simply depends on the fragility or robustness of the underlying monopartite graphs. Moreover, as for monopartite graphs, biased percolation can be mapped to the universality class of random removal on a different scale-free bipartite graph.

To obtain our results, we extended the generating functions theory for monopartite graphs to incorporate two types of nodes. Self-consistent equations for the generating functions allowed the determination of the critical fraction of the percolation process. Our results show that the critical fraction of a bipartite network can be written as a function of the critical fractions of the associated monopartite graphs. If at least one of those monopartite graphs is robust, also the bipartite network will be robust.  Based on these conclusions, we constructed a phase diagram in Fig.~1 picturing the different percolation regimes. Also the scaling of the critical threshold as a function of the network size was found to be an extension of the scaling of the critical fractions of the monopartite graphs. To calculate the critical exponents close to the percolation transition, a finite-size scaling theory was developed using the analogy between edge percolation and the $q\rightarrow1$ limit of the $q$-state Potts model. The scaling theory was solved by expanding the self-consistent equations close to the percolation threshold. Results for the exponents are given in Table 1. All exponents only depend on the constant $\alpha$ through the constants $\overline{\gamma}_A$ and $\overline{\gamma}_B$, defined as $\overline{\gamma}_i = (\gamma_i - \alpha)/(1-\alpha)$. The critical behavior of biased percolation with bias exponent $\alpha$ on a bipartite net with topological exponents $\gamma_i$ with $i=A,B$ thus is the same as the critical behavior of random removal on a network with exponents $\overline{\gamma}_i$. Therefore, we extended the main result of the study concerning monopartite graphs to bipartite graphs.

The generating functions theory we introduced can in principle be extended to include general multipartite networks with an arbitrary number of types of  nodes. In such networks, nodes can share edges with different types of nodes. For each type of nodes, a new generating function must be introduced, thereby increasing the complexity of the mathematics greatly. Since every new generating function requires an additional self-consistent equation, it becomes impossible to extract specific results, except in certain limiting cases. Much progress on this scheme has already been worked out in Ref.~\cite{allard2009}, but a calculation of critical exponents still remains an open issue. Further, in Ref.~\cite{moreira2} the authors performed a study of the extremum
events of scale-free networks, thereby focussing on the statistics of
the extreme connectivities. In this context, it would also be
interesting to investigate the evolution of such distribution functions
during the process of our network reconstruction and more especially
near the point of percolation.

\section*{Acknowledgements}
 H.H. is Research Assistant and B.V.S. is Post-Doctoral Researcher of the Fund for Scientific Research - Flanders (FWO-Vlaanderen). J.O.I. is immensely grateful to Nihat for his meticulous and spirited guidance throughout the ``polar liquid crystal period" (1984-1989), and, ever since he first met Nihat in 1979, for his scale-free hospitality, his appetite for re-entrant farce and his effervescently percolating friendship.


\begin{thebibliography}{250}
\bibitem{noot} Note that in Ref.~\cite{moreira} also the peripheral bias process with negative values of $\alpha$ is discussed. We will not focus on peripheral bias in the article at hand since this process is not well defined for the simultaneous approach used in this text~\cite{hooyberghs2009}.
\end{thebibliography}
\end{document}